\documentclass[12pt]{article}
\usepackage{latexsym} 
\usepackage{epsfig,epsf}
\usepackage{graphicx}
\usepackage{amsfonts}
\usepackage{amsmath}
\usepackage{amsthm}
\usepackage{amssymb}
\usepackage{cite}
\usepackage{cancel}

\def\subeqnarray{\arraycolsep1pt
    \def\@eqnnum\stepcounter##1{\stepcounter{subequation}%
        {\reset@font\rm(\theequation\alph{subequation})}}
\jot5mm     \eqnarray}

\makeatother

\def\be{\begin{equation}}
\def\lb#1{\label{#1}}
\def\ee{\end{equation}}
\def\bea{\begin{eqnarray}}
\def\eea{\end{eqnarray}}
\def\ba{\begin{array}}
\def\ea{\end{array}}
\def\dd{\partial}

\def\one#1{#1^{\raise5pt\hbox{$\scriptstyle\!\!\!\!1$}}\,{}}
\def\two#1{#1^{\raise5pt\hbox{$\scriptstyle\!\!\!\!2$}}\,{}}

\def\II{\hbox{{1}\kern-.25em\hbox{l}}}
\def\p#1{(\ref{#1})}
\makeatletter
\def\binrel@#1{\begingroup
  \setboxz@h{\thinmuskip0mu
    \medmuskip\m@ne mu\thickmuskip\@ne mu
    \setbox\tw@\hbox{$#1\m@th$}\kern-\wd\tw@
    ${}#1{}\m@th$}%
  \edef\@tempa{\endgroup\let\noexpand\binrel@@
    \ifdim\wdz@<\z@ \mathbin
    \else\ifdim\wdz@>\z@ \mathrel
    \else \relax\fi\fi}%
  \@tempa
}
\let\binrel@@\relax
\def\overset#1#2{\binrel@{#2}%
  \binrel@@{\mathop{\kern\z@#2}\limits^{#1}}}
\def\underset#1#2{\binrel@{#2}%
  \binrel@@{\mathop{\kern\z@#2}\limits_{#1}}}
\makeatother
\newfont{\bbd}{msbm10 scaled\magstep1}
\def\C{\hbox{\bbd C}}

\def\R{\hbox{\bbd R}}

\def\Z{\hbox{\bbd Z}}

\begin{document}


\begin{center}
{\LARGE \bf{On the $6j$-symbols for $\mathrm{SL}(2,\mathbb{C})$ group}}

\bigskip

{\large \sf S.\,E. Derkachov$^{a}$ 
 and V.\,P. Spiridonov$^b$ } 


\begin{itemize}
\item[$^a$]
{\it St. Petersburg Department of the Steklov Mathematical Institute
of Russian Academy of Sciences,
Fontanka 27, 191023 St. Petersburg, Russia}
\item[$^b$]
{\it Laboratory of Theoretical Physics, JINR, Dubna, 141980,  Russia  }
\end{itemize}
\end{center}


\begin{abstract}
We study $6j$-symbols, or Racah coefficients for tensor products of
infinite-dimensional unitary
principal series representations of the group $\mathrm{SL}(2,\mathbb{C})$.
These symbols were constructed earlier by Ismagilov and we rederive his
result (up to some slight difference associated with
equivalent representations) using the Feynman diagrams technique. The resulting
$6j$-symbols are expressed either as a triple integral over complex plane,
or as an infinite bilateral sum of integrals of the Mellin-Barnes type.
\end{abstract}

\smallskip
\begin{flushright}
\em To the memory of Ludwig Dmitrievich Faddeev
\end{flushright}


{\small \tableofcontents}

\section{Introduction}

The problem of decomposition of a tensor product of irreducible representations
of classical groups to the direct sum of such representations with
the help of $3nj$-symbols is a well known old subject of investigations.
Despite of very many results obtained in this field by Clebsch, Gordan, Wigner,
van der Waerden, Fock, Racah, Naimark, Biedenharn, and many other researches,
it is not completed yet and continues to be developed.
In a detailed investigation of the atomic spectra,
Racah \cite{racah} constructed a closed form expression
for $6j$-symbols of finite-dimensional representations of $\mathrm{SU}(2)$ group. It is
given by a terminating $_4F_3$ hypergeometric series and determines a set of classical
orthogonal polynomials called Racah polynomials \cite{aar}.
For an outline of the theory of $3nj$-symbols
and a list of relevant references, see the handbook \cite{VMK}.

The group $\mathrm{SL}(2,\mathbb{C})$ is one of the most important Lie groups, since it is
the smallest rank nonabelian group over the field of complex numbers $\mathbb C$
\cite{Gelfand-Naimark}. It coincides with the Lorentz group and therefore its
finite-dimensional representations play a crucial role in four-dimensional quantum
field theory, since they describe observable elementary particles.
Irreducible infinite-dimensional representations of $\mathrm{SL}(2,\mathbb{C})$
also have found appropriate applications in physics. They
emerge in a spin chain model that appears in the high-energy regime of
quantum chromodynamics, see~\cite{Lipatov,Lipatov:1993qn} and \cite{FK,Lipatov:1993yb}.
Therefore investigation of the representation
theory  of this group does not need additional justifications.

This paper is devoted to a consideration of $6j$-symbols, or Racah
coefficients (operators) for the tensor product of unitary infinite-dimensional
principal series representations of the $\mathrm{SL}(2,\mathbb{C})$ group.
The $3j$-symbols, or Clebsch-Gordan coefficients for such representations
have been constructed by Naimark long ago \cite{Naimark}.
They are defined by a single valued function of three complex variables,
describing the representation space, and depend on three integer and three real parameters.
The projectors onto irreducible components of the corresponding twofold tensor products are
given by integral operators with such kernel functions. Despite of the
importance of the problem of building $6j$-symbols for the
$\mathrm{SL}(2,\mathbb{C})$ group dealing with threefold tensor products,
for the unitary principal series representations they were constructed
only recently by Ismagilov in \cite{Ismag,Ismag2}.

The previous most close result on this subject was obtained in the work \cite{Groene},
where an integral transform related to the Wilson function was considered and
$6j$-symbols for the tensor products involving the unitary principal series
representation of the group $\mathrm{SU}(1,1)$ were constructed.
The results of Ismagilov open the final
chapter of the program of building $3nj$-symbols for the smallest rank groups.
As follows from the general $\mathrm{SL}(2,\mathbb{C})$ group representation
theory \cite{Gelfand-Naimark}, it remains to consider similar problems for the
cases involving the complementary series representation, as well as the
non-unitary representations.
In the present work we rederive the results of Ismagilov
using a different approach, namely the Feynman diagrams techniques,
and give two different types of integral representations for these $6j$-symbols.

The number of applications of $6j$-symbols is quite large ranging from
quantum mechanics, where they describe the angular momentum dynamics,
to quantum gravity, statistical mechanics, knot invariants, etc. For instance,
the operator intertwining equivalent principal series representations of the
$\mathrm{SL}(2,\mathbb{C})$ group (it is described in the next section) plays
a crucial role in the construction of general solutions of the vertex type
Yang-Baxter equation \cite{DM}. In a similar way, the $6j$-symbols
considered in this paper should define solutions of a different type
Yang-Baxter equation related to
IRF (``interaction round a face'') models in statistical mechanics.

In the last decades quantum deformations of the $sl(2)$ algebra
have been investigated from various points of view. In particular,
the modular double of $U_q(sl(2,\mathbb{R}))$ was introduced by Faddeev
in \cite{fad:mod} and $6j$-symbols for the unitary principal series
representation of this algebra have been constructed in \cite{PT}.
A further extension of these considerations to the simplest quantum
supergroup is given in \cite{PSS}. We expect that our results can
be lifted to the complex extension of these quantum groups as well
(see \cite{DM1,DMV} for related results).

The paper is organized as follows. In Sect. 2 we outline the structure of
$\mathrm{SL}(2,\mathbb{C})$ group and its principal series representation.
In Sect. 3 we describe the structure of Clebsch-Gordan coefficients
for the tensor product of two such representations and consider their
biorthogonality and completeness relations.  Sect. 4 contains main results
of our work --- a new derivation of the Racah coefficients for relevant
representations in the form of a kernel of an integral operator relating
different bases of threefold tensor products.
In Sect. 5 we provide a Mellin-Barnes representation for these
$6j$-symbols. In the Appendix we collected some handbook formulae and
an auxiliary material.

\section{$\mathrm{SL}(2,\mathbb{C})$ group}
\subsection{Representations of the group and the intertwining operator} \lb{SL2rep}

Let us describe some basic facts from the representation theory
of the group $\mathrm{SL}(2,\mathbb{C})$ \cite{Gelfand,Gelfand-Naimark}.
They are formulated in a form that
will be natural for dealing with the Racah coefficients and
corresponding projection operators.

Usually $\mathrm{SL}(2,\C)$ group representations are realized
in the space of single-valued functions $\Phi(z,\bar{z})$ on the complex
plane, $z\in\mathbb{C}$, with $\bar z$ being the complex conjugate of $z$.
The non-unitary principal series representation~\cite{Gelfand}
is parameterized by a pair of generic complex numbers $(s,\bar{s})$
subject to the single constraint $2\,(s-\bar{s}) \in \mathbb{Z}$.
We refer to them as {\it spins} in what follows. In order to avoid
misunderstanding we emphasize that $s$ and $\bar{s}$ are not complex
conjugates of each other. As usual, for a given matrix
$$
g = \left(\begin{array}{cc}
  \alpha & \beta \\
  \gamma & \delta \\
\end{array}\right) \in \mathrm{SL}(2,\C)
$$
one can consider $\mathrm{SL}(2,\C)$ group action on the two-dimensional plane
coordinates $x,y\in\mathbb{C}$ of the form
\begin{equation}\label{ism}
( x , y) \to  ( x , y) g
=(\alpha x +\gamma y, \beta x+\delta y),
\end{equation}
or
\begin{equation}\label{our}
\left({x \atop  y}\right) \to   g^{-1}
\left({x \atop  y}\right)
=\left({\delta x -\beta y \atop -\gamma x+\alpha y}\right).
\end{equation}
If the latter transformation is used then, after denoting $z=y/x$, one comes to
the representation $\mathrm{T}^{(s,\bar{s})}$ determined explicitly by
the corresponding linear fractional transformation~\cite{Gelfand}
\begin{equation}\label{tsl2}
\left[ \mathrm{T}^{(s,\bar{s})}(g)\,\Phi \right](z,\bar{z})
= \left(\delta -\beta z\right)^{2s}\,
\left(\bar{\delta} -\bar{\beta}\bar{z}\right)^{2\bar{s}}
\, \Phi\left(\frac{-\gamma+\alpha z}{\delta-\beta z} , \frac{-\bar{\gamma}+\bar{\alpha}
\bar{z}}{\bar{\delta}-\bar{\beta}\bar{z}}\right).\ \
\end{equation}

In \cite{Ismag2} Ismagilov used the first option \eqref{ism} which,
after denoting $z=x/y$, yields an equivalent though slightly
differently looking representation
\begin{equation}\label{tsl2ISM}
\left[ \mathrm{T}_{a}(g)\,\Phi \right](z,\bar{z}) =
\left(\beta z+\delta\right)^{a-1}\,
\left(\bar{\beta}\bar{z}+\bar{\delta}\right)^{\bar{a}-1}
\, \Phi\left(\frac{\alpha z+\gamma}{\beta z+\delta} , \frac{\bar{\alpha}
\bar{z}+\bar{\gamma}}{\bar{\beta}\bar{z}+\bar{\delta}}\right).
\end{equation}
We connect the representation parameters in \eqref{tsl2} and
\eqref{tsl2ISM} as $a=2s+1,\,  \bar{a}= 2\bar{s} +1$.
In \cite{Ismag2} the representations are taken to be
unitary principal series with the restrictions
$a: = m+i\sigma, \; \bar{a}: = - m+i\sigma,\; m\in\Z, \;  \sigma\in\mathbb{R}.$
In this case one has $a-\bar a=2s-2\bar{s}=2m \in 2\mathbb{Z}$, an even integer.

We start our considerations from the general non-unitary representation,
which assumes that the representation parameters have the form
\begin{equation}\label{rep}
a= \frac{ m}{2} + b +i\sigma, \quad \bar a= - \frac{m}{2}+ b
+ i\sigma,\quad  m\in\Z, \quad  b,\sigma\in\mathbb{R}.
\end{equation}
The unitary case corresponds to the choice $b=0$ and arbitrary integer
$m$. In \cite{Ismag2} the following function was used as a representation character
$$
\phi_k(a,z): = |z|^{i\sigma - m -k} z^{m}, \quad k\in\Z.
$$
Instead of this notation, we employ the following convention
$$
[z]^{a}: = z^{a} \bar{z}^{\bar{a}} =|z|^{2\bar{a}}z^{a-\bar a}
=|z|^{2b +2i\sigma - m} z^{m},
$$
which is a replacement of the function $\phi_0(2a,z)$ in  \cite{Ismag2}.

Taking the matrix $g$ lying in a vicinity of the unit matrix,
$g = 1 + \varepsilon\, \mathcal{E}_{ik}$,
where $\mathcal{E}_{ik}$ are traceless $2 \times 2$ matrices:
$
(\mathcal{E}_{ik})_{jl} =
\delta_{i j} \delta_{kl}-\frac{1}{2} \delta_{ik} \delta_{jl}\, ,
$
it is not difficult to find generators of the Lie algebra $s\ell(2,\mathbb{C})$
$\mathrm{E}_{ik}$ and $\bar{\mathrm{E}}_{ik}$,
$$
\mathrm{T}^{(s,\bar{s})}(1+\varepsilon\, \mathcal{E}_{ik})\,\Phi(z,\bar{z})=
\Phi(z,\bar{z})+\left(\varepsilon\,
\mathrm{E}_{ik}+\bar\varepsilon\,\bar{\mathrm{E}}_{ik}\right)\,
\Phi(z,\bar{z})+O(\varepsilon^2)\,.
$$
Explicitly, the generators $\mathrm{E}_{ik}$, $\bar{\mathrm{E}}_{ik}$ are
given by the first-order differential operators which we represent as
$2\times 2$ matrices $\mathrm{E}^{(s)}$ and $\bar{\mathrm{E}}^{(\bar{s})}$:
\begin{equation}\label{Egl2}
\mathrm{E}^{(s)} = \left(%
\begin{array}{cc}
  \mathrm{E}_{11} & \mathrm{E}_{21} \\
  \mathrm{E}_{12} & \mathrm{E}_{22} \\
\end{array}%
\right) = \left(%
\begin{array}{cc}
  z\partial-s & -\partial \\
  z^2\partial -2s\,z & -z\partial+s \\
\end{array}%
\right) = \begin{pmatrix}
1 & 0 \\ z & 1
\end{pmatrix}
\begin{pmatrix}
-s-1 & -\dd \\ 0 & s
\end{pmatrix}
\begin{pmatrix}
1 & 0 \\ -z & 1
\end{pmatrix},
\end{equation}
with the matrix $\mathrm{\bar{E}}^{(\bar{s})}$ (yielding the
generators $\mathrm{\bar{E}}_{ik}$ in a similar way) obtained from $\mathrm{E}^{(s)}$ after
the replacements $z \to \bar{z}\,, \partial \to \bar{\partial}$, and $s \to \bar{s}$.

It is well known that the representations characterized by the
parameters $s, \bar s$ and $-1-s, -1-\bar s$ (or $a, \bar a$ and
$-a, -\bar a$) are equivalent  (the values of Casimir operators for
them coincide) \cite{Gelfand-Naimark}.
There exists an integral operator $\mathrm{M}$ which intertwines such
equivalent principal series representations $\mathrm{T}^{(s,\bar{s})}$
and $\mathrm{T}^{(-1-s,-1-\bar{s})}$ for generic complex $s$ and $\bar{s}$,
\begin{equation}
\label{splet}
\mathrm{M}(s,\bar{s})\,\mathrm{T}^{(s,\bar{s})}(g) =
\mathrm{T}^{(-1-s,-1-\bar{s})}(g)\,\mathrm{M}(s,\bar{s})\,.
\end{equation}
Relations \eqref{splet} can be reformulated as a set of
intertwining relations for the Lie algebra generators
\begin{equation}
\label{spletE}
\mathrm{M}(s,\bar{s})\,\mathrm{E}^{(s)} =
\mathrm{E}^{(-1-s)}\,\mathrm{M}(s,\bar{s}),\qquad
\mathrm{M}(s,\bar{s})\,\mathrm{\bar{E}}^{(\bar{s})} =
\mathrm{\bar{E}}^{(-1-\bar{s})}\,\mathrm{M}(s,\bar{s}).
\end{equation}
This $\mathrm{M}$-operator can be written in the following form~\cite{Gelfand}
(for a justification of the taken normalization factor, see \cite{CDS,DM})
\be \lb{M}
\left[ \,\mathrm{M}(s,\bar{s}) \Phi\,\right](z,\bar{z}) =
\frac{i^{-|2s-2\bar s|}}{\pi}
\,\frac{\Gamma\left(s+\bar s+|s-\bar s|+2\right)}
{\Gamma\left( -s-\bar s+|s-\bar s| -1\right)}
\int \mathrm{d}^2 x\, \frac{\Phi(x,\bar{x})}{[z-x]^{2s+2}},
\ee
where $\Gamma(x)$ is the standard gamma function.
This is a well-defined operator for generic values of $s$ and $\bar{s}$.
Despite of the diverging integral for the discrete values $2s = n,\,
2\bar{s} = \bar{n}$, $n,\,\bar{n} \in \mathbb{Z}_{\geq 0}$,
it remains well defined in this case too due the appropriate
normalizing factor (see the next section).

The described intertwining operator has a meaning of the
pseudodifferential operator. Such an interpretation is reached
with the help of the following explicit Fourier transformation
\begin{equation}\label{A}
A(\alpha,\bar\alpha)\,\int d^2 z\, \frac{\mathrm{e}^{ipz+i\bar{p}\bar{z}}}
{z^{1+\alpha}\bar{z}^{1+\bar\alpha}} =
p^{\alpha}\bar{p}^{\bar\alpha}, \qquad  \alpha-\bar\alpha\in\Z,
\end{equation}
where the measure is defined as
$d^2 z=dxdy=\frac{i}{2}dzd\bar z$ (with $z=x+iy$, $\bar z=x-iy$)  and
the normalization constant has the canonical form~\cite{Gelfand}
\begin{equation}\label{Aconst}
A(\alpha,\bar\alpha):= \frac{i^{-|\alpha-\bar\alpha|}}{\pi}
\,\frac{\Gamma\left(\frac{\alpha+\bar\alpha+|\alpha-\bar\alpha|+2}{2}\right)}
{\Gamma\left(\frac{-\alpha-\bar\alpha+|\alpha-\bar\alpha|}{2}\right)}.
\end{equation}
For non-integer values of $\alpha$
the form of this constant can be simplified
\begin{equation}
A(\alpha,\bar\alpha) =
\frac{i^{\bar\alpha-\alpha}}{\pi  a(\alpha+1)},
\qquad a(\alpha):=\frac{\Gamma(1-\bar\alpha)}{\Gamma(\alpha)},
\label{Aa}\end{equation}
which can be checked by substitution of the relation $\alpha=\bar \alpha +m,\, m\in \mathbb{Z},$
for general non-unitary principal series representation and application of the reflection
formula for the gamma function.

Let us replace in formula \eqref{A} the complex variables $p$ and $\bar p$ by differential
operators, $p \to i\partial_x$ and $\bar{p} \to i\partial_{\bar{x}}$.
Then one can use the standard finite-difference operator
$\mathrm{e}^{a\partial_x}f(x)=f(x+a)$ in order to set by definition
\begin{eqnarray} \label{kern}  &&
\left(i\partial_z\right)^{\alpha}
\left(i\partial_{\bar{z}}\right)^{\bar\alpha}\Phi(z,\bar z) :=
A(\alpha,\bar\alpha)\,\int d^2 x\,
\frac{\Phi(x,\bar x)}{[z-x]^{1+\alpha}},
\\ && \makebox[0em]{}
[z-x]^{1+\alpha}:= (z-x)^{1+\alpha} (\bar{z}-\bar x)^{1+\bar{\alpha}}, \quad
\alpha - \bar{\alpha} \in \mathbb{Z}.
\lb{not1} \end{eqnarray}
The constraint on the exponents $\alpha,\bar{\alpha}$ in \p{not1} ensures that
the function $[z-x]^{\alpha}$ is single-valued.
If one takes the separate holomorphic part, then it has a branch cut,
but a special choice of the antiholomorphic multiplier yields the
single-valued function.
It is this pseudodifferential operator that is used for fixing the normalizing
factor in the intertwining operator \eqref{M} -- one simply sets
$\mathrm{M}(s,\bar{s}) := \left[i\partial_z\right]^{2s+1}$.
In particular, for $s=\bar s= -1/2$ one has $\mathrm{M}(s,\bar{s})= \II$
(the unit operator), see \cite{Gelfand}.

\subsection{Decoupling of the finite-dimensional representations}

The fact that finite-dimensional representations can be derived
from the general principal series representation by the reduction
is well known \cite{Gelfand-Naimark,CDS}. Indeed,
for the discrete set $2s = n,\, 2\bar{s} = \bar{n}$,
$n,\,\bar{n} \in \mathbb{Z}_{\geq 0}$, the integral operator \eqref{M}
becomes a finite order differential operator $\left(i\partial_z\right)^{n+1}
\left(i\partial_{\bar{z}}\right)^{\bar{n}+1}$. This follows not from
the formal identification $\mathrm{M}(s,\bar{s})= \left[i\partial_z\right]^{2s+1}$,
but from the rigorous consideration of singularities of meromorphic functions of $s$
appearing after the action of this operator
on sufficiently smooth functions $\Phi(z,\bar{z})$ \eqref{M}
and careful consideration of the limits $2s \to n,\, 2\bar{s} \to \bar{n}$,
see in \cite{Gelfand} a description of the tempered distribution $z^{-n-2}{\bar z}^{-\bar n-2}$.
The main transformation law (\ref{tsl2}) implies that for such discrete values of spins
an $(n+1)(\bar{n}+1)$-dimensional representation decouples from the
general infinite-dimensional one. This is evident,
since the $(n+1) (\bar{n}+1)$-dimensional vector space
spanned by polynomials $z^{k} \bar{z}^{\bar{k}}$,
$k = 0, 1 , \cdots , n$ and $\bar{k} = 0, 1 ,\cdots , \bar{n}$,
is invariant with respect to the action of the operators
$\mathrm{T}^{(s,\bar{s})}(g)$.

This picture is nicely captured by the intertwining operator $\mathrm{M}(s,\bar{s})$ \eqref{M}.
One can introduce a single generating function for these basis polynomials
$[z-x]^{n} = (z-x)^{n} (\bar{z} - \bar{x})^{\bar{n}}$,
where $x$, $\bar{x}$ are some auxiliary variables. Clearly the series
expansion of $[z-x]^{n}$ in $x$ and $\bar{x}$ yields needed
vectors $z^{k} \bar{z}^{\bar{k}}$,$k = 0, 1 , \cdots , n$,
$\bar{k} = 0, 1 , \cdots , \bar{n}$.
Then from  relation (\ref{splet}) it follows that
the space annihilated by the operator $\mathrm{M}(s,\bar{s})$ (the null-space)
is invariant under the action of the operators
$\mathrm{T}^{(s,\bar{s})}(g)$. Therefore  any nontrivial invariant
null-space yields a sub-representation.  For $2s =n$ and $2\bar{s} = \bar{n}$,
the intertwining operator turns into the differential
operator $\dd^{n+1} \bar{\dd}^{\bar{n}+1}$ which annihilates
the generating function $[z-x]^{n}$.
However, the full null-space includes all harmonic functions, i.e.
it is much bigger.
The image of the intertwining operator $\mathrm{M}(-1-s,-1-\bar{s})$ is
invariant under the action of $\mathrm{T}^{(s,\bar{s})}(g)$, which follows
from formula \eqref{splet}. We have also the relation
\be \lb{dint1}
\left[ \,\mathrm{M}(-1-s,-1-\bar{s})
\Phi\,\right](z,\bar{z}) =
\frac{i^{-|2s-2\bar s|}}{\pi}
\,\frac{\Gamma\left(-s-\bar s+|s-\bar s|\right)}
{\Gamma\left( s+\bar s+|s-\bar s| +1\right)}
\int d^2 x\, (z-x)^{2s}(\bar{z}-\bar{x})^{2\bar{s}}\,
\Phi(x,\bar{x}).
\ee
After dropping the numerical factor $\Gamma\left(-s-\bar s+|s-\bar s|\right)$
diverging for $2s = n$ and $2\bar{s} = \bar{n}$,
we clearly see that the image of $\mathrm{M}(-1-s,-1-\bar{s})$
is a polynomial of $z$ and $\bar z$ forming the needed finite-dimensional subspace.
So, the  polynomial finite-dimensional subspace is formed
as an intersection of the null-space of $\mathrm{M}(s,\bar{s})$
and the image of the properly normalized operator $\mathrm{M}(-1-s,-1-\bar{s})$
with the spins $2s = n$ and $2\bar{s} = \bar{n}$.

The fact that the intertwining operator annihilates the generating
function $[z-x]^{n}$ can be established using the inversion
property of the intertwining operator. Indeed, the notation
$\mathrm{M}(s,\bar{s}) = \left[i\partial_z\right]^{2s+1}$ and
$\mathrm{M}(-1-s,-1-\bar{s}) = \left[i\partial_z\right]^{-1-2s}$
formally suggests that $\mathrm{M}(s,\bar{s})\,\mathrm{M}(-1-s,-1-\bar{s}) = \II$.
However, this relation cannot be true for positive integer values of the spins.
Let us rewrite this inversion relation after substituting
the explicit forms of the kernels for integral operators
$\mathrm{M}(-1-s,-1-\bar{s})$ \eqref{dint1} and $\II$ (given by the Dirac delta-function)
\begin{equation} \label{delta0}
\left[i\partial_z\right]^{2s+1} [z-x]^{2s} =
\pi i^{|2s-2\bar s|}\frac{\Gamma\left( s+\bar s+|s-\bar s| +1\right)}
{\Gamma\left(-s-\bar s+|s-\bar s|\right)}
\,\delta^2(z-x).
\end{equation}
For $2s\to n,\, 2\bar{s}\to \bar{n},\, n,\bar n\in\Z_{\geq 0},$ the multiplier
in the denominator $\Gamma\left(-s-\bar s+|s-\bar s|\right)$ acquires the poles.
For $n\geq \bar n$ it is $\Gamma(-\bar n)$ and
for  $n\leq \bar n$ it is $\Gamma(- n)$,
so that the right-hand side of this relation vanishes. Therefore,
$
\left[i\partial_z\right]^{n+1} [z-x]^{n} = 0, \; n, \bar n=0,1,2,\ldots,
$
i.e. the generating function of the finite-dimensional
representations $[z-x]^{n}$ is the kernel function of
the properly normalized operator $\mathrm{M}(-1-n/2,-1-\bar n/2)$.

\section{Decomposition of the tensor product of two representations}

Decomposition of the tensor product of two principal
series representations to irreducible components
has been constructed by Naimark~\cite{Naimark}. The projection operator
$$
\mathrm{T}_{a_1}\otimes \mathrm{T}_{a_2} \xrightarrow{\mathrm{P}(a_1,a_2|a_3)} \mathrm{T}_{a_3},
$$
is given by the following integral operator
$$
\Phi(z_1,z_2) \xrightarrow{\mathrm{P}(a_1,a_2|a_3)}
\left[\mathrm{P}(a_1,a_2|a_3)\,\Phi\right](z_3) =
\int \mathrm{d}^2\,z_1\mathrm{d}^2\,z_2\,
W\left(
{a_1,  a_2, a_3 \atop z_1, z_2, z_3}
\right)\, \Phi(z_1,z_2)\,.
$$
The kernel function represents the Clebsch-Gordan coefficients and has
the following exlpicit form
\begin{align}\label{W}
W\left(
{a_1,  a_2, a_3 \atop z_1, z_2, z_3}
\right) = [z_2-z_1]^{-\frac{1+a_1+a_2+a_3}{2}}
[z_3-z_1]^{-\frac{1+a_1-a_2-a_3}{2}}
[z_2-z_3]^{-\frac{1-a_1+a_2-a_3}{2}},
\end{align}
which coincides with the expression given by Naimark in \cite{Naimark}
after the identification $z_3=z$.
This function is fixed up to an overall normalization constant by the
requirement of covariance
\begin{eqnarray}\nonumber &&
\left[\beta z_3+\delta\right]^{a_3-1}
\left[\mathrm{P}(a_1,a_2|a_3)\,\Phi\right]
\left(\frac{\alpha z_3+\gamma}{\beta z_3+\delta}\right)
\\  && \makebox[0.5em]{}
= \int \mathrm{d}^2\,z_1\mathrm{d}^2\,z_2\,
W\left(
{a_1,  a_2, a_3 \atop z_1, z_2, z_3}
\right)\,
\left[\beta z_1+\delta\right]^{a_1-1}\,
\left[\beta z_2+\delta\right]^{a_2-1}
\Phi\left(\frac{\alpha z_1+\gamma}{\beta z_1+\delta},
\frac{\alpha z_2+\gamma}{\beta z_2+\delta}\right).
\label{covar} \end{eqnarray}

The $3j$-symbols \eqref{W} were derived in \cite{Naimark} for unitary principal series
representations. However, we stress that the corresponding derivation does not
depend on whether the parameter $i\sigma$ is purely imaginary or a general complex number.
Therefore formulas \eqref{W}, \eqref{covar} are true for general non-unitary principal
series representation with $b\neq 0$ in the parametrization of spin variables
\eqref{rep}. It should be noticed that in formula \eqref{W} one has the exponent
of the form $\alpha:=\frac{1+a_1+a_2+a_3}{2}$ and similar ones which do not
preserve the general restriction on spin values $\alpha-\bar \alpha\in\Z$. As pointed out
in \cite{Naimark}, this means that the nontrivial Clebsch-Gordan coefficients
exist only in the cases when the integers $m_1,m_2,m_3$ entering the definition
of parameters $a_1,a_2,a_3$ satisfy the constraint that $m_1+m_2+m_3$ is an
even integer (in \cite{Ismag,Ismag2} this condition was resolved by forcing
all $m_j$ to be even integers).

In an infinitesimal form the global relation \eqref{covar} is
equivalent to the system of defining equations
\begin{align}\label{covar1}
\left(\mathrm{E}^{(-a_1)}_{z_1}+\mathrm{E}^{(-a_2)}_{z_2}+
\mathrm{E}^{(a_3)}_{z_3}\right)W\left(
{a_1,  a_2, a_3 \atop z_1, z_2, z_3}
\right) = 0.
\end{align}
Here for brevity we use the superscript $a_j = 2s_j+1$ for labeling the generators
instead of the previously used $s_j$-variables, in terms of which equivalent
representations are described by the reflection $s_j\to-s_j-1$.

\begin{figure}[t]
\centerline{\includegraphics[width=0.4\linewidth]{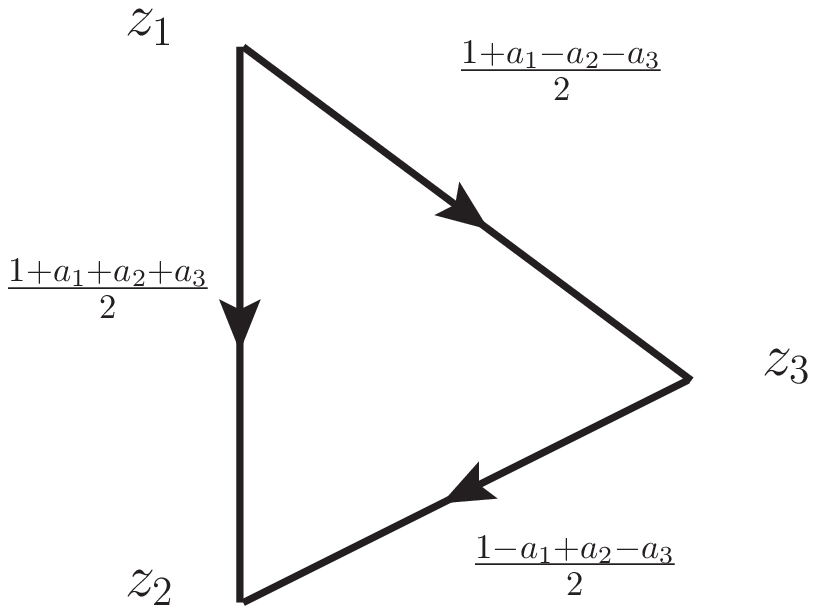}}
\caption{Diagrammatic representation of $W\left({a_1, a_2, a_3 \atop z_1,z_2,z_3}\right)$.}
\label{W1}
\end{figure}

Now we present the basic elements of the diagram technique
which will be used throughout the paper.
The kernels of integral operators  are represented in the form of
two-dimensional Feynman diagrams. The propagator is given by the following expression
\begin{equation}
\frac{1}{[z-w]^\alpha}\equiv\frac{1}{(z-w)^\alpha (\bar z-\bar w)^{\bar\alpha}}=
\frac{(\bar z-\bar w)^{\alpha-\bar\alpha}}{|z-w|^{2\alpha}}=\frac{(-1)^{\alpha-\bar\alpha}}{[w-z]^{\alpha}},
\end{equation}
where $\alpha-\bar\alpha$ is an integer. It is depicted on the diagrams by the
lint with the arrow directed from
point $w$ to $z$ with the index $\alpha$ corresponding to scaling exponents.
 The diagrammatic representation for our main building block
$W\left({a_1, a_2, a_3 \atop z_1,z_2,z_3}\right)$ is given by Fig.~\ref{W1}.
We have the diagram with three external vertices and due to the required behavior under
$\mathrm{SL}(2,\mathbb{C})$-transformations~(\ref{covar}) this diagram coincides up to an
overall coefficient with the simple conformal triangle.
The name ``conformal triangle" is due to the fact that the system of
equations~(\ref{covar1}) coincides with the set of Ward identities
for the three-point conformal invariant Green function in two-dimensional
conformal field theory \cite{BPZ}.

For the unitary principal series representation, which corresponds to
the choice $b=0$ in \eqref{rep},
the complex conjugation is equivalent to the change of signs of all spin variables:
\begin{align}\label{barW}
\overline{W\left(
{a_1,  a_2, a_3 \atop z_1, z_2, z_3}\right)} =
W\left(
{-a_1,  -a_2, -a_3 \atop z_1, z_2, z_3}\right).
\end{align}
From now on we shall assume that
the function $\overline W$ \eqref{barW} is a complex conjugate of $W$.
The representations with the parameters $a$ and $-a$ are known to be
equivalent. Therefore function \eqref{barW} is the Clebsch-Gordan coefficient
for the decomposition problem when all three involved representations are
replaced by the equivalent ones.

The kernel of the dual projection operator
$
\mathrm{T}_{a_3} \xrightarrow{\mathrm{P}(a_3|a_1,a_2)} \mathrm{T}_{a_1}\otimes \mathrm{T}_{a_2},
$
is determined precisely by the function \eqref{barW}
$$
\Phi(z_3)\xrightarrow{\mathrm{P}(a_3|a_1,a_2)}
\left[\mathrm{P}(a_3|a_1,a_2)\,\Phi\right](z_1,z_2) =
\int \mathrm{d}^2\,z_3\,W\left(
{-a_1, -a_2,-a_3 \atop z_1,z_2,z_3}\right)\, \Phi(z_3)\,.
$$
This follows from the biorthogonality relation considered in the next section.

\subsection{Orthogonality and completeness}

\begin{figure}[t]
\centerline{\includegraphics[width=0.9\linewidth]{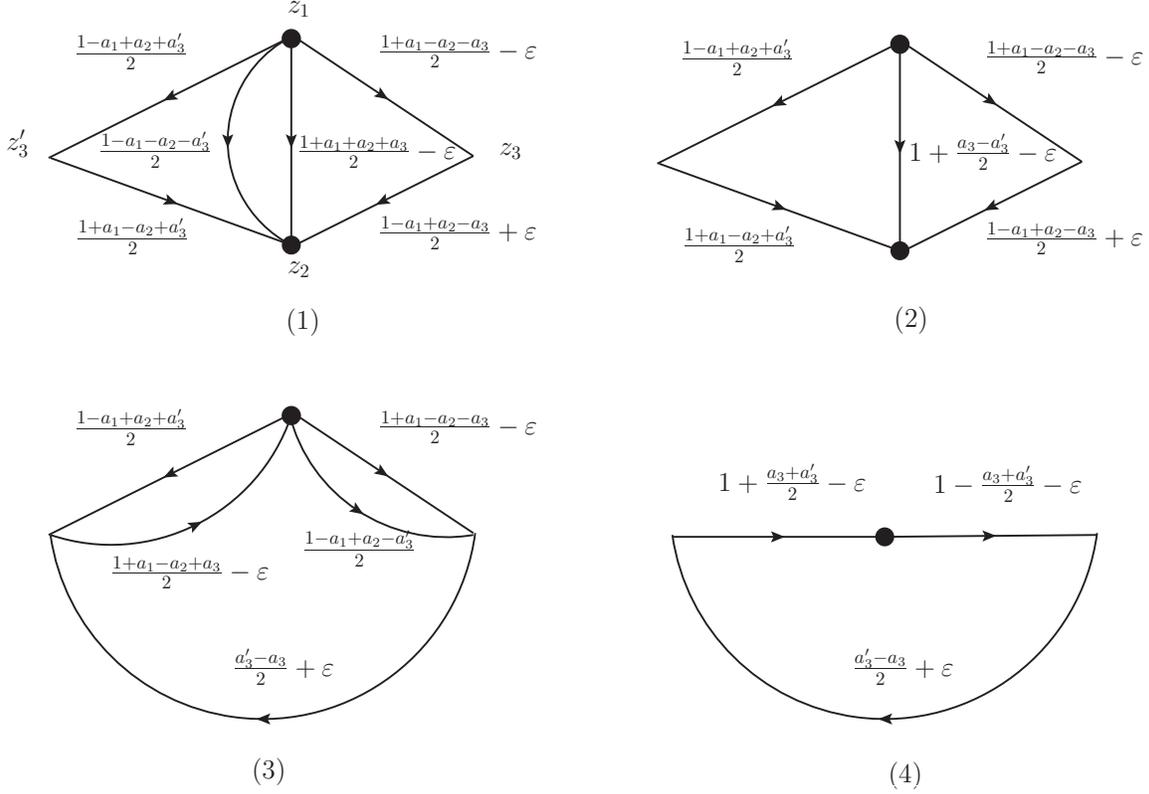}}
\caption{(1) The initial regularized orthogonality diagram.
(2) The diagram after replacement of the middle double line by a single line.
(3) The star-triangle transformation at the lower vertex.
(4) The same diagram (up to a sign) after replacement of all double
lines by single ones. The final integration can be done using the
chain relation~(\ref{Chain})}
\label{ort0}
\end{figure}

Let us prove the following (bi)orthogonality relation for unitary principal
series representation (to which we are limiting from now on)
\begin{eqnarray} \nonumber &&
\int \mathrm{d}^2\,z_1\mathrm{d}^2\,z_2\,
W\left(
{-a_1, -a_2,-a_3^{\prime} \atop z_1,z_2,z_3^{\prime}}\right)
W\left(
{a_1, a_2,a_3 \atop z_1,z_2,z_3}\right)
\\  && \makebox[2em]{}
=\rho^{-1}(a_3)\delta_R(a_3-a_3^{\prime})\,\delta^2(z_3-z_3^{\prime}) + B(a_1,a_2,a_3)\,\frac{\delta_R(a_3+a_3^{\prime})}{[z_3-z_3^{\prime}]^{1-a_3}},
\label{ort}\end{eqnarray}
where $\rho(a_3)$ and $B(a_1,a_2,a_3)$ are some weight functions.
The parametric delta-function $\delta_R(a-a')$ has the form
$$
\delta_R(a-a')=\delta_{m\, m'}\delta\left(\sigma-\sigma'\right), \quad
a=\frac{m}{2}+i\sigma,\quad a'=\frac{m'}{2}+i\sigma'.
$$

Emergence of the second term in~(\ref{ort}) is a direct consequence of the fact
that two representations $\mathrm{T}_a$ and $\mathrm{T}_{-a}$ are equivalent
and that there exists an intertwining operator  with the kernel
$[z-z^{\prime}]^{-1-a}$. Namely, one has the equality
\begin{align}\label{intW}
\int \mathrm{d}^2\,z_3^{\prime}\,\frac{1}{[z_3-z_3^{\prime}]^{1+a_3}}
W\left(
{a_1, a_2,a_3 \atop z_1,z_2,z_3^{\prime}}\right) = A(a_1,a_2,a_3)\,
W\left({a_1, a_2,-a_3 \atop z_1,z_2,z_3}\right),
\end{align}
which is equivalent to the star-triangle relation~(\ref{Star})
and can be easily checked to have
\begin{equation}\label{A'}
A(a_1,a_2,a_3)=\pi \frac{a\left(\frac{1+a_1-a_2-a_3}{2},1+a_3\right)}
{a\left(\frac{1+a_1-a_2+a_3}{2}\right)},
\end{equation}
where $a(\alpha,\beta,\ldots):=a(\alpha)a(\beta)\ldots$
and the function $a(\alpha)$ was defined in \eqref{Aa}.
Application of the relation~(\ref{intW}) to (\ref{ort}) shows
an inevitability of the second term on the right-hand side
corresponding to the Clebsch-Gordan coefficient with the change $a_3\to -a_3$.
It leads also to some relations between the functions
$\rho(a_3)$, $A(a_1,a_2,a_3)$ and $B(a_1,a_2,a_3)$
which can be used as a crosscheck of the final results.

The second term $\propto \delta(a_3+a_3')$
was explicitly presented by Lipatov in \cite{Lipatov} in the special case
$a_1=a_2=0$ (namely, for the notation
$W\left({0,\; 0,\; a\atop z_1,z_2,z_0}\right)=
E^{{n,\nu}}(z_j)/|z_2-z_1|^2$ with $a=n+2i\nu,\, n\in\mathbb{Z}$).
In the case of quantum
groups, appearance of such a term in the corresponding orthogonality relation
was considered in \cite{DF} and \cite{HPS}.

In Fig.~\ref{ort0} we show a step-by-step calculation procedure of the
diagram corresponding to the left-hand side of the orthogonality relation
where blobs in the vertices denote integrations over the corresponding coordinates.
To avoid an ill-defined integral expression, we introduce an
$\varepsilon$-regularization. Namely, we replace the coefficient $W$ for
the right-hand side triangle by the expression $W_\varepsilon$,
which differs from the original $W$ by addition of an infinitesimally
small real number $\varepsilon= \bar \varepsilon>0$
to the line indices, as indicated in Fig.~\ref{ort0}. Note that the sign of
$\varepsilon$ is firmly fixed by the demand of convergence of the
emerging Feynman integrals. Indeed, the upper right diagram contains
the line contributing to the integral over $z_1$ (or $z_2$)
the factor $1/[z_2-z_1]^{1+(a_3-a_3')/2-\varepsilon}$. The singularity
at the point $z_1=z_2$ must be integrable. Therefore, for the
values of parameters $a_3\approx a_3'$ relevant for the orthogonality
relation (see below), one must have $\varepsilon>0$
(the integral $\int d^2z/[z]^\gamma$ converges near $z=0$ for Re$(\gamma)<1$).
Similarly, the last diagram integral over $z_1$ contains the factor
$1/[z_1-z_3']^{1+(a_3+a_3')/2-\varepsilon}$ and a similar one with $z_1$
replaced by $z_2$. Again, for the domain of value of parameters
of interest $a_3+a_3'\approx 0$ the integral converges for $\varepsilon>0$.

Performing carefully all four steps of the
computation procedure one should change several times the line directions
with the accompanying change of signs, $[x-y]^a=(-1)^{a-\bar a}[y-x]^a$.
So, a transition to the right bottom diagram yields the
additional multiplier $(-1)^{\frac{m_1-m_2-m_3'}{2}}$,
and the change of direction of the lower line in the latter diagram yields
the multiplier $(-1)^{\frac{m_3'-m_3}{2}}$.
Collecting all emerging factors together, we obtain the expression
\begin{eqnarray}\nonumber  &&
\int \mathrm{d}^2\,z_1\mathrm{d}^2\,z_2\,
W\left(
{-a_1, -a_2,-a_3^{\prime} \atop z_1,z_2,z_3^{\prime}}\right)
W_{\varepsilon}\left(
{a_1, a_2,a_3 \atop z_1,z_2,z_3}\right)
\\  \nonumber  && \makebox[2em]{}
= (-1)^{\frac{m_1-m_2-m_3}{2}}\pi^2\,a\Big(\frac{1-a_1+a_2-a_3}{2}+\varepsilon,\,
\frac{1+a_1-a_2+a_3^{\prime}}{2},
1+\frac{a_3-a_3^{\prime}}{2}-\varepsilon,
\\  && \makebox[3em]{}
1+\frac{a_3+a_3^{\prime}}{2}-\varepsilon,
1-\frac{a_3+a_3^{\prime}}{2}-\varepsilon,  2\varepsilon\Big)\,
\frac{1}{[z_3-z_3^{\prime}]^{1+\frac{a_3^{\prime}-a_3}{2}-\varepsilon}}.
\label{regort}\end{eqnarray}

Now we can carefully investigate what happens in the limit $\varepsilon \to 0$.
First of all, we note that in the generic situation, when
$a_3+a_3^{\prime}\neq 0$ and $a_3-a_3^{\prime}\neq 0$, everything is
regular in $\varepsilon$ and, due to the presence of the function
$a(2\varepsilon)\sim \varepsilon$, the whole expression vanishes
in the limit $\varepsilon \to 0$.

When $a_3+a_3^{\prime} \approx 0$, which happens for $m_3=-m_3'$
and $\sigma_3\approx - \sigma_3^{\prime}$, we have
\begin{eqnarray*} &&
\lim_{\varepsilon \to 0} a\left(1+\frac{a_3+a_3^{\prime}}{2}-\varepsilon,
1-\frac{a_3+a_3^{\prime}}{2}-\varepsilon, 2\varepsilon\right)
\\ && \makebox[1em]{}
=\delta_{m_3,-m_3'}\lim_{\varepsilon \to 0}
\frac{2\varepsilon}{\left(\frac{\sigma_3}{2}
+\frac{\sigma_3'}{2}\right)^2+\varepsilon^2}
= 4\pi \delta_{m_3,-m_3'}\delta\left(\sigma_3+\sigma_3^{\prime}\right).
\end{eqnarray*}
Here we use the relation $\varepsilon/(x^2+\varepsilon^2)\to \pi\delta(x)$
for $\varepsilon\to 0^+$ and $x\in\mathbb{R}$.
Note that, if we would take $\varepsilon<0,$ we would obtain
on the right-hand side a different sign.

As a result, on the right-hand side of equality \eqref{regort} we have
for $\varepsilon \to 0$
\begin{align}
(-1)^{\frac{m_1-m_2-m_3}{2}}4\pi^3
a\left(\frac{1-a_1+a_2-a_3}{2},\frac{1+a_1-a_2-a_3}{2},1+a_3\right)
\frac{\delta_R(a_3+a_3^{\prime})}
{[z_3-z_3^{\prime}]^{1-a_3}}.
\end{align}

Suppose now that $a_3-a_3^{\prime} \approx 0$, i.e. $m_3=m_3'$
and $\sigma_3\approx \sigma_3^{\prime}$. Then we use another formula
producing the delta-function
\begin{align}\label{deltafunction}
\lim_{\varepsilon \to 0}
\frac{\varepsilon}{[z_3-z_3^{\prime}]^{1-\varepsilon}}
= \pi \delta^2(z_3-z_3^{\prime})
\end{align}
valid for arbitrary complex $\varepsilon$. It emerges in the relation
\begin{eqnarray*} &&
\lim_{\varepsilon \to 0} \frac{\Gamma(-\frac{a_3-a_3^{\prime}}{2}+\varepsilon)}
{\Gamma(2\varepsilon)}\frac{1}{[z_3-z_3']^{1+\frac{a_3'-a_3}{2}
-\varepsilon}}
=\delta_{m_3,m_3'}\lim_{\varepsilon \to 0}
\frac{2\varepsilon}{\left(\frac{\sigma_3}{2}
-\frac{\sigma_3'}{2}\right)^2+\varepsilon^2}
\frac{\frac{a_3-a_3^{\prime}}{2}+\varepsilon}
{[z_3-z_3']^{1+\frac{a_3'-a_3}{2}-\varepsilon}}
\\ && \makebox[4em]{}
= 4\pi^2 \delta_{m_3,m_3'}\delta\left(\sigma_3-\sigma_3^{\prime}\right)
\delta^2(z_3-z_3').
\end{eqnarray*}
As a result, for $a_3-a_3^{\prime} \approx 0$ we find
$$
(-1)^{\frac{m_1-m_2-m_3}{2}}4\pi^4 a\left(\frac{1-a_1+a_2-a_3}{2},
\frac{1+a_1-a_2+a_3}{2},1+a_3,
1-a_3\right)\delta_R(a_3-a_3^{\prime})\delta^2(z_3-z_3^{\prime}).
$$

Applying now the reflection formulas for $a(\alpha)$-function given in the Appendix
and collecting all the terms together, we obtain relation \eqref{ort} with
\begin{eqnarray}\label{answ1} &&
\rho(a_3) =-\frac{a_3\bar{a}_3}{4\pi^4}, \qquad
B(a_1,a_2,a_3) = 4\pi^3 \frac{a\left(\frac{1-a_1+a_2-a_3}{2}, 1+a_3\right)}
{a\left(\frac{1-a_1+a_2+a_3}{2}\right)}.
\end{eqnarray}
Note that $\rho(a)$ is a positively defined weight function, since $a\bar{a}=-(m^2/4+\sigma^2)$.

The completeness relation for $3j$-symbols of interest was established by Naimark in \cite{Naimark}
(see there formulae (114) and (115), as well as Theorem 3).
Its form depends on the parity of the integer parameters defining the representations.
Let us fix $a_j=\frac{m_j}{2}+i\sigma_j$, $m_j\in\mathbb{Z},\, \sigma_j\in\mathbb{R}$, $j=1,2,3$,
and denote for brevity $a_3\equiv a=\frac{m}{2}+i\sigma$.

Suppose that $m_1+m_2$ is an even integer. Then one has the following completeness relation
\begin{align}\label{completness}
\sum_{m\in 2\mathbb{Z}}\int_{\mathbb{R}}\mathrm{d}\sigma \int_{\mathbb{C}}\mathrm{d}^2z \,
\frac{\rho(a)}{2}\,
W\left({-a_1, -a_2,-a  \atop z_3,z_4,z}\right)W\left(
{a_1, a_2,a  \atop z_1,z_2,z}\right)
=\delta^2(z_1-z_3)\,\delta^2(z_2-z_4).
\end{align}
As a cross check of this equality,
let us multiply it by $W\left({-a_1, -a_2,-a' \atop z_1,z_2,z'}\right)$ and integrate over $z_1$ and $z_2$.
Applying the orthogonality relation \eqref{ort} and integrating out the corresponding
delta functions, we come to the trivial identity due to the equality
$\rho(a)A(-a_1,-a_2,-a)B(a_1,a_2,a)=(-1)^m,$ where $A, B, $ and $\rho$ are fixed
in \eqref{ort} and \eqref{A'}.

Assume now that $m_1+m_2$ is an odd integer. In this case one can write
\begin{align}\label{completness2}
\sum_{m\in 2\mathbb{Z}+1}\int_{\mathbb{R}}\mathrm{d}\sigma \int_{\mathbb{C}}\mathrm{d}^2z \,
\frac{\rho(a)}{2}\, W\left({-a_1, -a_2,-a  \atop z_3,z_4,z}\right)W\left(
{a_1, a_2,a  \atop z_1,z_2,z}\right)
=\delta^2(z_1-z_3)\,\delta^2(z_2-z_4).
\end{align}

\section{Triple tensor products and the Racah coefficients}

Take now the tensor product of three representations $\mathrm{T}_{a_1}\otimes \mathrm{T}_{a_2}\otimes \mathrm{T}_{a_3}$ and decompose it to the sum of irreducible representations. This can be
done in two ways. The first possibility is
\begin{equation}
\mathrm{T}_{a_1}\otimes \mathrm{T}_{a_2} \otimes \mathrm{T}_{a_3}\xrightarrow{\mathrm{P}(a_1,a_2|c)}
\mathrm{T}_{c}\otimes \mathrm{T}_{a_3}\xrightarrow{\mathrm{P}(c,a_3|\ell)} \mathrm{T}_{\ell},
\label{first_dec}\end{equation}
which is realized by the integral operator
\begin{eqnarray} \nonumber &&
\Phi(z_1,z_2,z_3) \xrightarrow{\mathrm{P}(c,a_3|\ell)\mathrm{P}(a_1,a_2|c)}
\left[\mathrm{P}(c,a_3|\ell)\mathrm{P}(a_1,a_2|c)\,\Phi\right](z)
\\  && \makebox[2em]{}
= \int \mathrm{d}^2z_1\,\mathrm{d}^2z_2\,\mathrm{d}^2z_3\int  \mathrm{d}^2z_0\,
W\left({a_1, a_2,c \atop z_1,z_2,z_0} \right)
W\left({c, a_3,\ell \atop z_0,z_3,z}\right)\, \Phi(z_1,z_2,z_3).
 \nonumber \end{eqnarray}
Let us remind that the integer variables entering the representation parameters $a_j,c,\ell$
must satisfy the conditions that $m_1+m_2+m_c$ and $m_c+m_3+m_\ell$ are even integers.

\begin{figure}[t]
\centerline{\includegraphics[width=1.0\linewidth]{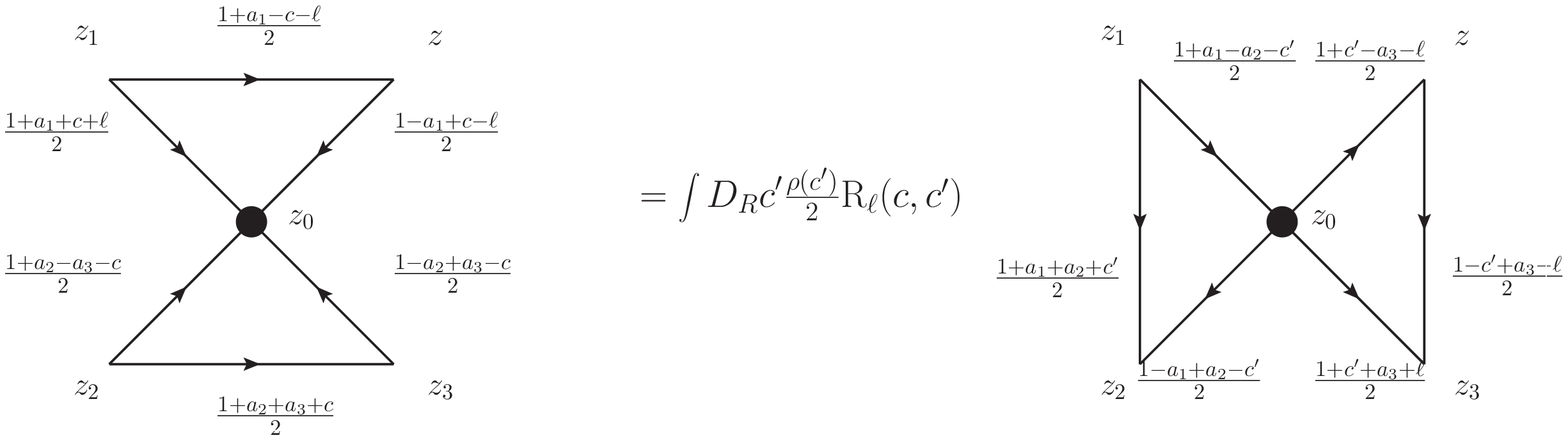}}
\caption{Diagrammatic representation of relation~(\ref{main}).}
\label{Main}
\end{figure}

The second possibility of decomposition
$$
\mathrm{T}_{a_1}\otimes \mathrm{T}_{a_2} \otimes \mathrm{T}_{a_3}\xrightarrow{\mathrm{P}(a_2,a_3|c)}
\mathrm{T}_{a_1}\otimes \mathrm{T}_{c}
\xrightarrow{\mathrm{P}(a_1,c|\ell)} \mathrm{T}_{\ell},
$$
is realized by another integral operator
\begin{eqnarray} \nonumber &&
\Phi(z_1,z_2,z_3) \xrightarrow{\mathrm{P}(a_1,c|\ell)\mathrm{P}(a_2,a_3|c)}
\left[\mathrm{P}(a_1,c|\ell)\mathrm{P}(a_2,a_3|c)\,\Phi\right](z)
\\  && \makebox[2em]{}
= \int \mathrm{d}^2z_1\,\mathrm{d}^2z_2\,\mathrm{d}^2z_3
\int \mathrm{d}^2z_0\,W\left(
{a_2,  a_3, c \atop z_2, z_3, z_0}
\right)W\left(
{a_1,  c, \ell \atop z_1, z_0, z}
\right)\, \Phi(z_1,z_2,z_3).
 \nonumber \end{eqnarray}
The $6j$-symbols, or Racah coefficients $\mathrm{R}_{\ell}$ are defined as
the kernel of the integral operator connecting these two decompositions
\begin{align}
\mathrm{P}(a_1,c|\ell)\mathrm{P}(a_2,a_3|c) =
\int \mathrm{D}_R c^{\prime}\, \frac{\rho(c^{\prime})}{2}\,\mathrm{R}_{\ell}(c,c^{\prime})\,
\mathrm{P}(c^{\prime},a_3|\ell)\mathrm{P}(a_1,a_2|c^{\prime}).
\end{align}
Explicitly, they are defined by the following integral equation
\begin{eqnarray} \nonumber &&
\int \mathrm{d}^2z_0\,W\left({a_2, a_3,c \atop z_2,z_3,z_0}
\right)W\left({a_1, c,\ell \atop z_1,z_0,z}\right)
 \\  && \makebox[2em]{}
 = \int \mathrm{D}_R c^{\prime}\,\frac{\rho(c^{\prime})}{2}\,\mathrm{R}_{\ell}(c,c^{\prime})\,
\int \mathrm{d}^2z_0\,W\left({a_1, a_2,c^{\prime} \atop z_1,z_2,z_0}
\right)W\left({c^{\prime}, a_3,\ell \atop z_0,z_3,z}\right).
 \label{main} \end{eqnarray}
Here we set $c'=m/2+i\sigma$ and define the measure $\int\mathrm{D}_R c'$
either as $\sum_{m\in2\mathbb{Z}}\int_{\mathbb{R}}\mathrm{d}\sigma$
or  $\sum_{m\in 2\mathbb{Z}+1}\int_{\mathbb{R}}\mathrm{d}\sigma$
depending on whether $m_1+m_2$ is even or odd, respectively.
The diagrammatic representation of this relation is given in Fig.~\ref{Main}.

\begin{figure}[t]
\centerline{\includegraphics[width=0.8\linewidth]{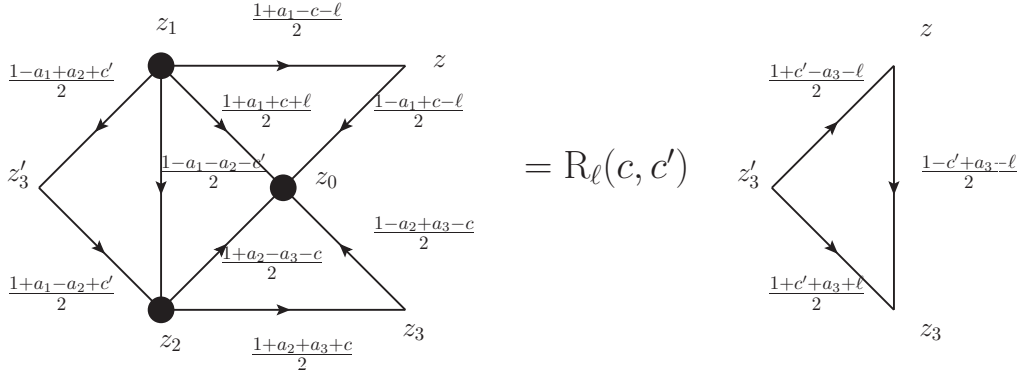}}
\caption{Diagrammatic representation of relation~(\ref{main1}).}
\label{Main1}
\end{figure}

The expression for the kernel $\mathrm{R}_{\ell}(c,c^{\prime})$
can be obtained by using orthogonality relation~(\ref{ort})
\begin{align}\label{main1} \makebox[-1em]{}
\int \mathrm{d}^2z_0\,\mathrm{d}^2z_1\,\mathrm{d}^2z_2\,
W\left({-a_1,-a_2,-c^{\prime} \atop z_1,z_2,z_3^{\prime}}\right)
W\left({a_2, a_3,c \atop z_2,z_3,z_0}\right)
W\left({a_1, c,\ell \atop z_1,z_0,z}\right) = \mathrm{R}_{\ell}(c,c^{\prime})
W\left({c^{\prime}, a_3,\ell \atop z_3^{\prime},z_3,z}\right).
\end{align}
The diagrammatic representation of this equality is given in Fig.~\ref{Main1}.
We have a diagram with three external vertices $z, z_3, z_3'$ and, due to the conformal invariance,
it should coincide with the conformal triangle up to an overall coefficient $\mathrm{R}_{\ell}(c,c^{\prime})$.
Vice versa, equations \eqref{main} and \eqref{main1} can be derived from the completeness relations
\eqref{completness}, \eqref{completness2}.

In the asymptotic regime when one of the coordinates
$z, z_3, z_3'$ goes to infinity, e.g.
$$
W\left(
{a_1, a_2,a_3 \atop z_1,z_2,z_3}
\right) \xrightarrow{z_1\to \infty} [z_1]^{-a_1}
[z_2-z_3]^{-\frac{1-a_1+a_2-a_3}{2}},
$$
we can reduce this three-point diagram to the two-point one.
So, for $z_3\to \infty$ we obtain
\begin{align}\label{main3}
\int \frac{\mathrm{d}^2z_0\,\mathrm{d}^2z_1\,\mathrm{d}^2z_2}
{[z_0-z_2]^{\frac{1+a_2-a_3-c}{2}}}\,
\overline{W\left(
{a_1, a_2,c^{\prime} \atop z_1,z_2,z_3^{\prime}}\right)}
W\left(
{a_1, c,\ell \atop z_1,z_0,z}
\right) = \frac{\mathrm{R}_{\ell}(c,c^{\prime})}
{[z-z_3^{\prime}]^{\frac{1+c^{\prime}-a_3-\ell}{2}}}.
\end{align}
\begin{figure}[t]
\centerline{\includegraphics[width=0.8\linewidth]{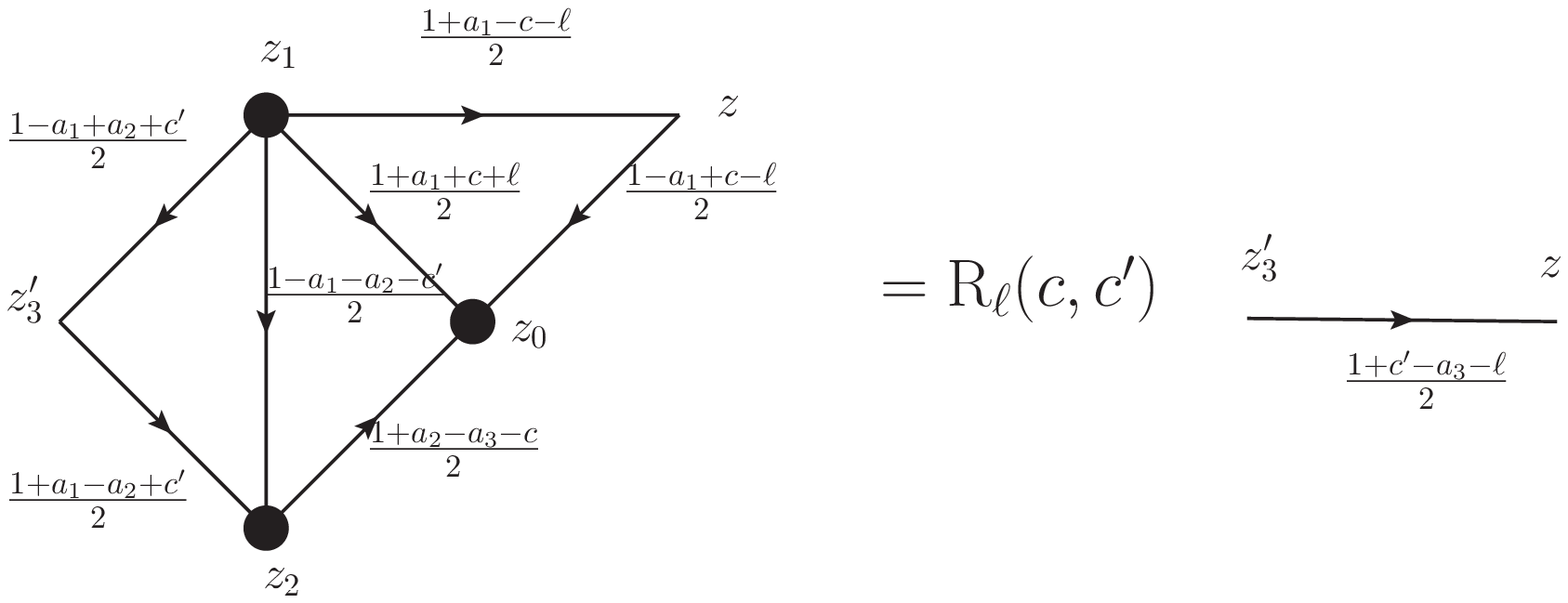}}
\caption{Diagrammatic representation of relation~(\ref{main3}).}
\label{Main3}
\end{figure}
For $z_3^{\prime}\to \infty$, we have
\begin{align}\label{main4}
\int
\frac{\mathrm{d}^2z_0\,\mathrm{d}^2z_1\,\mathrm{d}^2z_2}
{[z_2-z_1]^{\frac{1-a_1-a_2-c^{\prime}}{2}}}\,
W\left(
{a_2,  a_3, c \atop z_2, z_3, z_0}
\right)W\left(
{a_1,  c, \ell \atop z_1, z_0, z}
\right) = \frac{\mathrm{R}_{\ell}(c,c^{\prime})}
{[z_3-z]^{\frac{1-c^{\prime}+a_3-\ell}{2}}},
\end{align}
\begin{figure}[t]
\centerline{\includegraphics[width=0.7\linewidth]{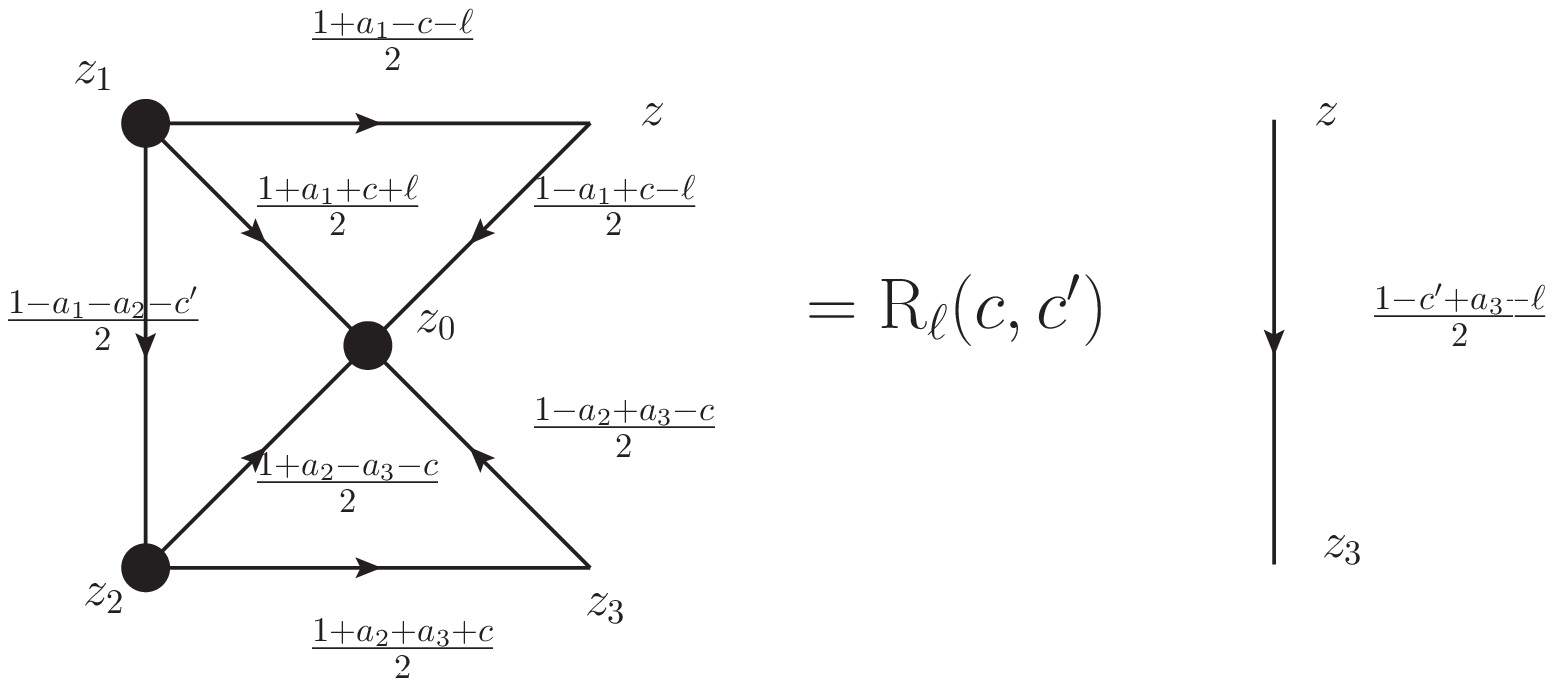}}
\caption{Diagrammatic representation of relation~(\ref{main4}).},
\label{Main4}
\end{figure}
and for $z\to \infty$
\begin{align}\label{main2}
\int \frac{\mathrm{d}^2z_0\,\mathrm{d}^2z_1\,\mathrm{d}^2z_2}
{[z_0-z_1]^{\frac{1+a_1+c+\ell}{2}}}
\,
\overline{W\left(
{a_1,  a_2, c^{\prime} \atop z_1, z_2, z_3^{\prime}}\right)}W\left(
{a_2,  a_3, c \atop z_2, z_3, z_0}
\right) = \frac{\mathrm{R}_{\ell}(c,c^{\prime})}
{[z_3-z_3^{\prime}]^{\frac{1+c^{\prime}+a_3+\ell}{2}}}.
\end{align}
\begin{figure}[t]
\centerline{\includegraphics[width=0.8\linewidth]{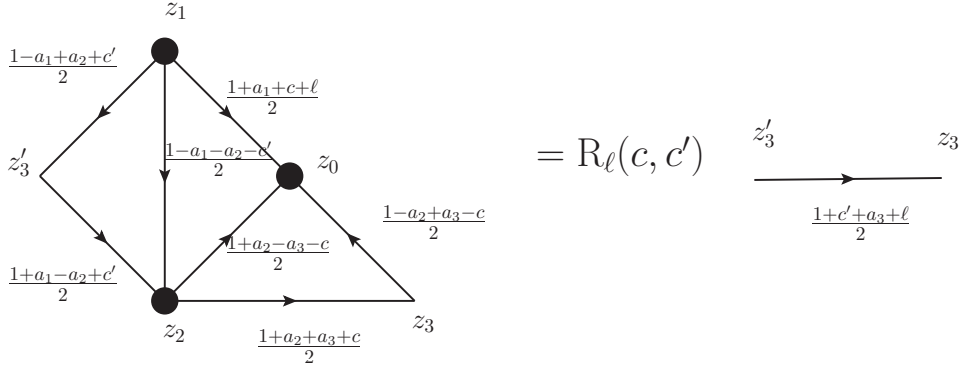}}
\caption{Diagrammatic representation of relation~(\ref{main2}).}
\label{Main2}
\end{figure}
These relations are depicted in Figs. \ref{Main3}--\ref{Main2}, where the limits
diagrammatically correspond to the removal of two lines with the
ends in the considered external vertex.
Resulting two-point diagrams are fixed again by the conformal invariance to be given
by a free propagator up an overall coefficient  $\mathrm{R}_{\ell}(c,c^{\prime})$,
which we call the value of the Feynman diagram of interest.

\begin{figure}[t]
\centerline{\includegraphics[width=0.8\linewidth]{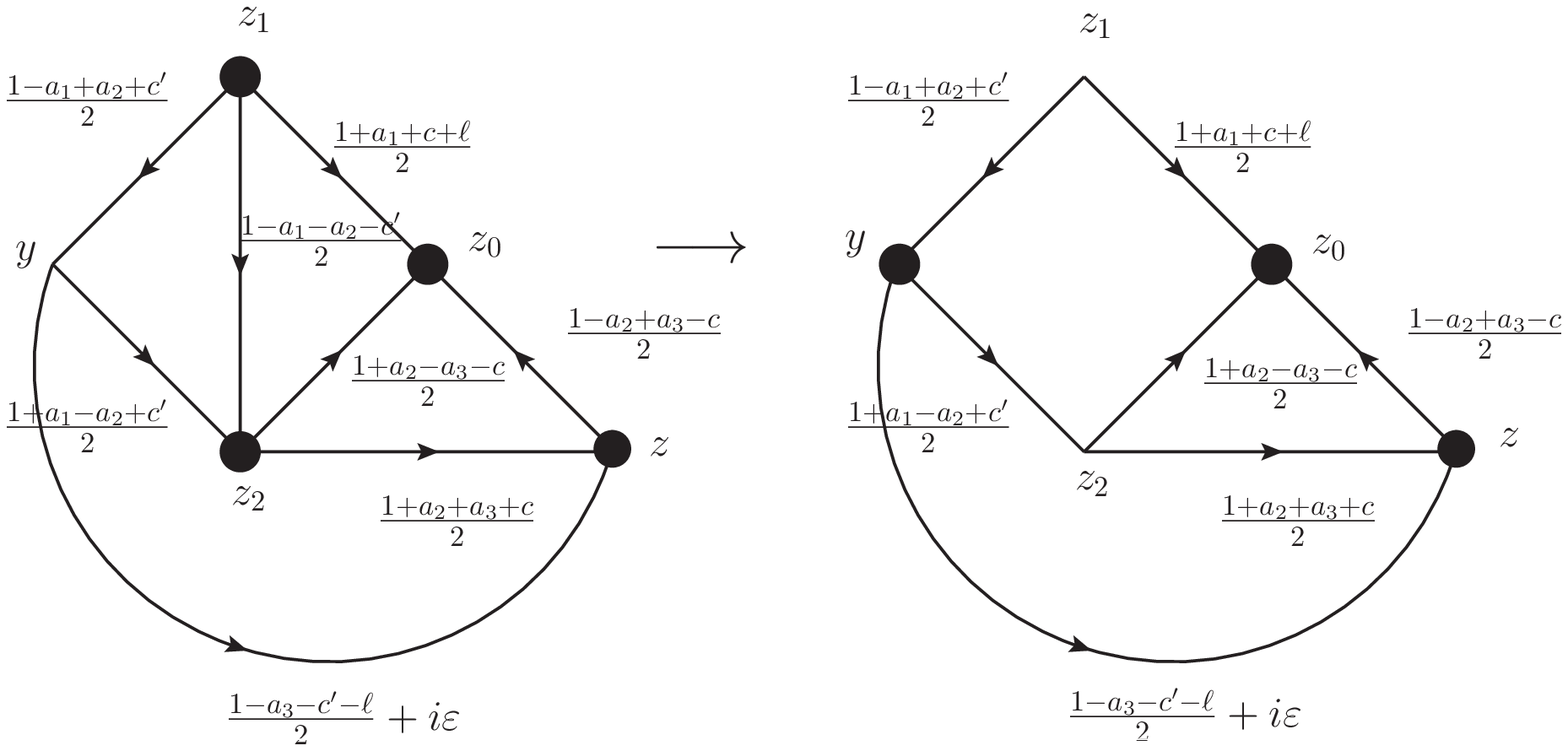}}
\caption{The parent diagram and the transition to an equivalent diagram.}
\label{FGI}
\end{figure}

All these three diagrams give identical evaluations, i.e. they represent symmetries of the
corresponding function $\mathrm{R}_{\ell}(c,c^{\prime})$ as a function of its parameters.
The origin of such a symmetry was discovered in~\cite{Gor-Isa}. Let us consider for example
the diagram in Fig.~\ref{Main2}. It belongs to the family of diagrams generated
by the parent diagram shown on the left-hand side of Fig.~\ref{FGI}. Namely, it emerges from it
after removing the line with the index $i\varepsilon+(1-a_3-c^{\prime}-\ell)/2$, $\varepsilon\in\mathbb{R}$.
The parent diagram has four integration vertices and one external vertex
and it equals to
\begin{equation}
\mathrm{R}_{\ell}(c,c^{\prime})\int \mathrm{d}^2 z
\frac{1}{[z-y]^{\frac{1+c^{\prime}+a_3+\ell}{2}}}\frac{1}
{[z-y]^{i\varepsilon+\frac{1-a_3-c^{\prime}-\ell}{2}}}
=\mathrm{R}_{\ell}(c,c^{\prime})2\pi^2 \delta(\varepsilon).
\label{parent}\end{equation}

The main observation of~\cite{Gor-Isa} is that all diagrams obtained from such parent
vacuum diagrams by removing one arbitrary line have the same value.
We have shown in Fig.~\ref{FGI} the transition to one of the possible equivalent diagrams by
removing the line with the index $(1-a_1-a_2-c^{\prime})/2$.
The resulting diagram contains only three integration vertices and has the value
$$
\mathrm{R}_{\ell}(c,c^{\prime})[z_1-z_2]^{-\frac{1+a_1+a_2+c^{\prime}}{2}-i\varepsilon}.
$$
If we multiply this expression by the removed line propagator $[z_1-z_2]^{\frac{a_1+a_2+c^{\prime}-1}{2}}$
and integrate over $z_1$, we get again relation \eqref{parent}.
Now we set in the right-hand side diagram in Fig.~\ref{FGI} $z_1=1,\, z_2=0$.
As a result, the propagator part disappears. Since $\mathrm{R}_{\ell}(c,c^{\prime})$
does not depend on $\varepsilon$, we can take the limit $\varepsilon\to 0$ in the
propagator connecting vertices $z$ and $y$. This yields an exact integral representation for
$\mathrm{R}_{\ell}(c,c^{\prime})$:
\begin{equation}
\mathrm{R}_{\ell}(c,c^{\prime}) = \int \mathrm{d}^2z\,
\Phi_2(a_1,a_2,a_3|\ell,c,z)
\overline{\Phi_1(a_1,a_2,a_3|\ell,c^{\prime},z)},
\label{R1}\end{equation}
where
\begin{eqnarray}\label{Phi1} &&
\overline{\Phi_1(a_1,a_2,a_3|\ell,c^{\prime},z)} =
\int \frac{ \mathrm{d}^2y }
{[y-1]^{\frac{1-a_1+a_2+c^{\prime}}{2}}
[-y]^{\frac{1+a_1-a_2+c^{\prime}}{2}}
[z-y]^{\frac{1-a_3-\ell+c^{\prime}}{2}}},
\\ && \makebox[-2em]{}
\Phi_2(a_1,a_2,a_3|\ell,c,z) = \frac{1}{[z]^{\frac{1+a_2+a_3+c}{2}}}
\int \frac{ \mathrm{d}^2z_0}
{[z_0-1]^{\frac{1+a_1+\ell+c}{2}}
[-z_0]^{\frac{1+a_2-a_3-c}{2}}
[z_1-z]^{\frac{1-a_2+a_3-c}{2}}}.
\label{Phi2}\end{eqnarray}

\begin{figure}[t]
\centerline{\includegraphics[width=0.9\linewidth]{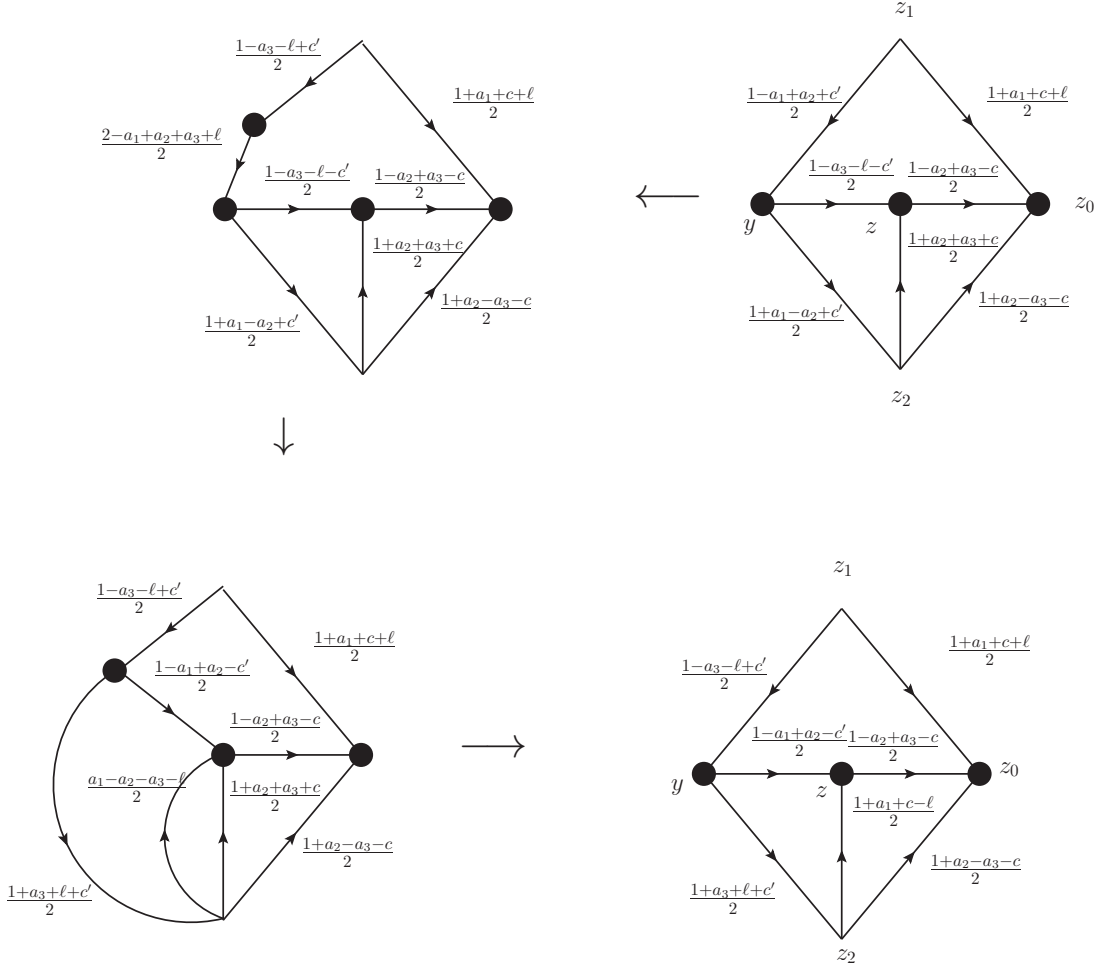}}
\caption{Transformation of the diagram.}
\label{ISM}
\end{figure}

In \cite{Ismag} Ismagilov has found the following representation for the same function
\begin{equation}\label{ismfin}
\mathrm{R}_{\ell}(c,c^{\prime}) = \int \mathrm{d}^2z\,
\Psi_2(a_1,a_2,a_3|\ell,c,z)
\overline{\Psi_1(a_1,a_2,a_3|\ell,c^{\prime},z)},
\end{equation}
where
$$
\overline{\Psi_1(a_1,a_2,a_3|\ell,c^{\prime},z)} =
\int \frac{\mathrm{d}^2y}
{[1-y]^{\frac{1-\ell-c^{\prime}-a_3}{2}}
[y]^{\frac{1+\ell-c^{\prime}+a_3}{2}}
[z-y]^{\frac{1-a_1+a_2+c^{\prime}}{2}}},
$$
$$
\Psi_2(a_1,a_2,a_3|\ell,c,z) =\frac{1}{[z]^{\frac{1+a_1-\ell+c}{2}}}
\int \frac{\mathrm{d}^2z_0}
{[1-z_0]^{\frac{1+a_1+\ell+c}{2}}
[z_0]^{\frac{1+a_2-a_3-c}{2}}
[z-z_0]^{\frac{1-a_2+a_3-c}{2}}}.
$$
We see that this expression has the same structure as our result, but indices of some
propagators are different (sign differences in
the arguments of $[1-y],[y], [1-z_0],[z_0]$ are inessential due to the constraints on the
parity of integers $m_j, m_c, m_{c'}, m_\ell$).
However, there is a symmetry transformation relating two expressions.
The corresponding chain of transformations of diagrams is shown in Fig.~\ref{ISM}.
In its right-upper corner we give a more compact form of our diagram in Fig.~\ref{FGI}.
Then we use the chain integration rule and the star-triangle relation from Fig.~\ref{Chain+Star}.
After that we arrive to the diagram in the right-lower corner in Fig.~\ref{ISM}.
Writing the corresponding integral representation one can see that it coincides
with Ismagilov's expression~(\ref{ismfin}) after the replacement of his
parameter $c^{\prime}$
by $-c^{\prime}$. So, our results almost coincide. This change $c^{\prime} \to -c^{\prime}$
corresponds to the replacement of the representation $\mathrm{T}_{c}$ in the first decomposition
\eqref{first_dec} by the equivalent representation $\mathrm{T}_{-c}$. Since this
is a nontrivial action, it is necessary to understand the source of such a difference
of our result with the one in \cite{Ismag}.

\section{Mellin-Barnes representation}

In this section we derive a Mellin-Barnes type representation
for the Racah coefficients described in the previous section.
Let us fix $s=(n+i\nu)/2$, $\bar s= (-n+i\nu)/2$.
There is a well-known representation for the two-dimensional delta function
\begin{align}\label{delta1}
\int D s \left[\frac{x}{y}\right]^{s}\,= (2\pi)^2\,[x]\,\delta^2(x-y),
\quad \int D s:= \sum_{n\in \mathbb{Z}}\int_{\nu\in\mathbb{R}}d \nu.
\end{align}

Take a real variable $\varepsilon=\bar\varepsilon>0$, which will serve as a regularization parameter
for infrared divergences. Then, with the help of formula \eqref{delta1} we can write
\begin{eqnarray}\nonumber &&
\frac{|y|^{2\varepsilon}}{[z-y]^{\alpha}} = \int d^2 x
\frac{|x|^{2\varepsilon}}{[z-x]^{\alpha}}\,\delta^2(x-y) = \frac{1}{(2\pi)^2}
\int D s \int d^2 x
\frac{1}{[z-x]^{\alpha}[x]^{1-s-\varepsilon}}\,
\frac{1}{[y]^{s}}
\\ && \makebox[6em]{}
=\frac{\pi a(\alpha)}{(2\pi)^2} \int D s\,
\frac{a(1-s-\varepsilon,1+s+\varepsilon-\alpha)}{[-z]^{\alpha-s-\varepsilon}[y]^{s}},
\nonumber \end{eqnarray}
where we used the chain integration rule~(\ref{Chain}). Denote $\alpha=(n_\alpha+i\nu_\alpha)/2$,
$\bar\alpha=(-n_\alpha+i\nu_\alpha)/2$, $n_\alpha\in\Z,\, \nu_\alpha\in\R$.
Poles of the integrand lie on the vertical half-lines at the points
$$
\nu=i (-n+2\varepsilon+2\Z_{\geq0}),\quad \nu_\alpha+i (n-n_\alpha+2\varepsilon-2\Z_{\geq0}).
$$

It is clearly seen that for $0<\varepsilon<1/2$ there are no singularities lying on the integration contour Im$(\nu)=0$ for all admissible values of $n\in\Z$.
Therefore we can change the integration contour to any contour lying in the strip
$\text{Im}(\nu)\in ]0,-1[$. After that we can take the limit $\varepsilon \to 0$
and come to the following Mellin-Barnes type
representation of the propagator with an arbitrary index $\alpha$,
\begin{align}\label{MB}
\frac{1}{[z-y]^{\alpha}} = \frac{1}{4\pi a(1-\alpha)}
\sum_{n\in\mathbb{Z}}\int_{L}d\nu\,
\frac{a(1-s,1+s-\alpha)}{[z]^{\alpha-s}[-y]^{s}},
\end{align}
where $L$ can be any contour lying in the strip $\text{Im}(\nu)\in ]0,-1[$.

\begin{figure}[t]
\centerline{\includegraphics[width=0.9\linewidth]{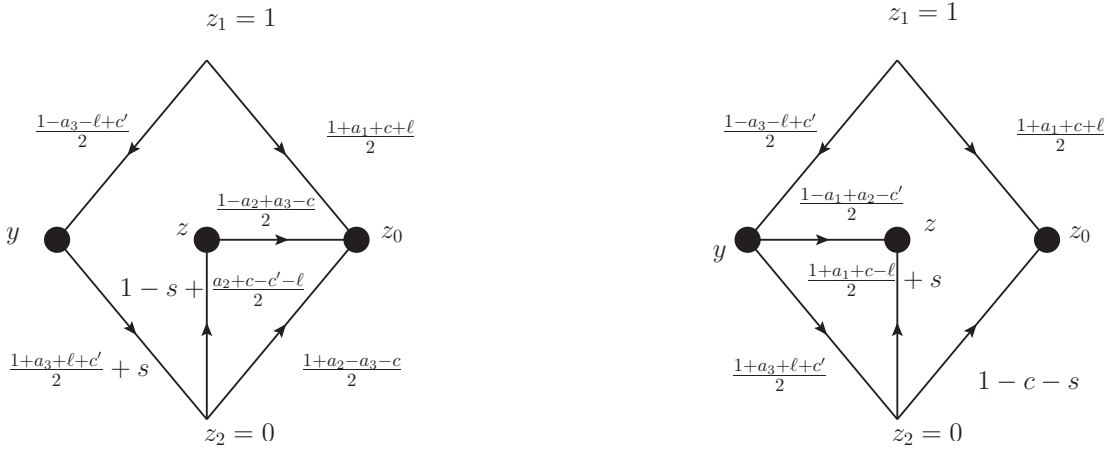}}
\caption{Mellin-Barnes representation diagrams.}
\label{ISM2}
\end{figure}

Now we apply this formula to the line connecting the points $y$ and $z$
in the last diagram of Fig. \ref{ISM}. This yields an ``integral'' over
the variable $s$ of the diagram given on the left-hand side of
Fig. \ref{ISM2}. However, the latter diagram can be calculated explicitly
with the help of the chain rule~(\ref{Chain}). Omitting the details of
computation, we obtain the following final Mellin-Barnes type
representation for the $6j$-symbols
\begin{eqnarray} \nonumber &&
\mathrm{R}_{\ell}(c,c^{\prime}) =
(-1)^{c^{\prime}-\bar c^{\prime}}\,\frac{\pi^2}{4}\frac{a\left(
\frac{1-a_3-\ell+c^{\prime}}{2},
\frac{1+a_1+c+\ell}{2}\right)}
{a\left(\frac{1+a_1-a_2+c^{\prime}}{2},
\frac{1+a_2-a_3+c}{2}\right)}
\\ &&  \makebox[2em]{} \times
\sum_{n\in\mathbb{Z}}\int_{L}d\nu\,
 \frac{a\left(\frac{1+a_1-a_2+c^{\prime}}{2}+s,
\frac{1-a_1-a_2+c^{\prime}}{2}+s,
\frac{1+a_3+\ell+c^{\prime}}{2}+s,\frac{1-a_3+\ell+c^{\prime}}{2}+s\right)}
{a\left(s,c^{\prime}+s,\frac{c^{\prime}+\ell-a_2-c}{2}+s,
\frac{c+c^{\prime}+\ell-a_2}{2}+s\right)}.
 \label{MB1} \end{eqnarray}
One can check that the integrands in \eqref{MB1} and in the $S$-function
entering the Mellin-Barnes representation given in Theorem 2 of
\cite{Ismag2} coincide after shifting the variable $s\to s+(a_2-a_1-c')/2$
(i.e. appropriate shifts of the summation variable $n$ and
integration variable $\nu$) and denoting $z=i\nu/2$.
However, the prefactor in front of the $S$-function
in \cite{Ismag2} misses the numerical multiplier $\pi^2/2i$ and
differs from ours by the replacement $c'\to -c'$, as before.

Equivalently, it is possible to apply formula~(\ref{MB})
to the line connecting $z$ and $z_0$ in the last diagram of Fig. \ref{ISM}.
In this way we come to the right-hand side diagram in Fig. \ref{ISM2}
which can be calculated again by using the rule~(\ref{Chain}).
This yields the second Mellin-Barnes type representation of interest
\begin{eqnarray} \nonumber  &&
\mathrm{R}_{\ell}(c,c^{\prime}) =
(-1)^{c^{\prime}-\bar c^{\prime}}\frac{\pi^2}{4}\,\frac{a\left(
\frac{1-a_3-\ell+c^{\prime}}{2},
\frac{1+a_1+c+\ell}{2}\right)}
{a\left(\frac{1+a_1-a_2+c^{\prime}}{2},
\frac{1+a_2-a_3+c}{2}\right)}
\\ && \makebox[2em]{} \times
\sum_{n\in\mathbb{Z}}\int_{L}d\nu\,
 \frac{a\left(\frac{1+a_2-a_3+c}{2}+s,
\frac{1+a_1-\ell+c}{2}+s,
\frac{1+a_2+a_3+c}{2}+s,\frac{1-a_1-\ell+c}{2}+s\right)}
{a\left(s,c+s,
\frac{a_2+c-\ell-c^{\prime}}{2}+s,
\frac{a_2+c-\ell+c^{\prime}}{2}+s\right)}.
\label{MB2} \end{eqnarray}
One can see that this expression is obtained from \eqref{MB1} simply by the
shift of the variable $s\to s+(a_2-\ell-c'+c)/2$.

\section{Conclusion}

In this work we have computed $6j$-symbols for the unitary principal series representaion
of the group $\mathrm{SL}(2,\mathbb{C})$, which are described by formulas \eqref{R1}-\eqref{Phi2}.
They coincide with the Racah coefficients obtained by Ismagilov \cite{Ismag} up to the
replacement in his expression for them the representation parameter $c'$ by $-c'$.
Note, however, that our result is slightly more general than in \cite{Ismag},
since we do not assume that the integer representation parameters $m_j$ are even.

As shown in \cite{Ismag}, the Mellin-Barnes representation for $\mathrm{R}_{\ell}(c,c^{\prime})$
can be rewritten in an equivalent form as a sum of the products of two $_4F_3$ hypergeometric
series with different arguments. We do not present the corresponding cumbersome expressions here.

We expect that the derived function $\mathrm{R}_{\ell}(c,c^{\prime}) $
describes Boltzmann weights of an IRF type integrable two-dimensional
statistical mechanics model and, so, solves the corresponding
Yang-Baxter equation. On the basis of the described approach it is
possible to build more general $6j$-symbols related
to the very-well poised $_9F_8$-series. In principle, following the results
of \cite{DMV}, it is possible to establish a relation to elliptic $6j$-symbols
described by the $V$-function presented in \cite{UMNsurvey} (an elliptic
extension of the Euler-Gauss hypergeometric function), which is
a subject for a separate consideration.

\smallskip
{\bf Acknowledgements.}
This work is supported by the Russian Science Foundation (project no. 14-11-00598).


\section{Appendix}

\begin{figure}[t]
\centerline{\includegraphics[width=0.6\linewidth]{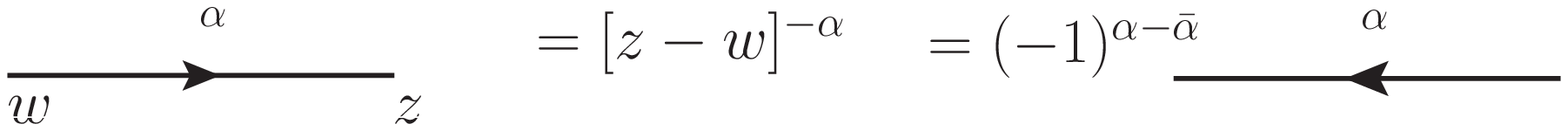}}
\caption{The propagator and a flip of the arrow.}
\label{line}
\end{figure}

In the diagram technique we use, the kernels of operators  are represented in the form of
two-dimensional Feynman integrals. The propagator, which is shown by the arrow directed from $w$ to $z$ and
index $\alpha$ attached to it as in Fig.~\ref{line}, is given by the following expression
\begin{equation}
\frac{1}{[z-w]^\alpha}\equiv\frac{1}{(z-w)^\alpha (\bar z-\bar w)^{\bar\alpha}}=
\frac{(\bar z-\bar w)^{\alpha-\bar\alpha}}{|z-w|^{2\alpha}}=\frac{(-1)^{\alpha-\bar\alpha}}{[w-z]^{\alpha}},
\end{equation}
where $\alpha-\bar\alpha=n_\alpha$ is an integer. On the same figure we indicate the
result of the flipping of the direction of the line.

\begin{figure}[t]
\centerline{\includegraphics[width=0.7\linewidth]{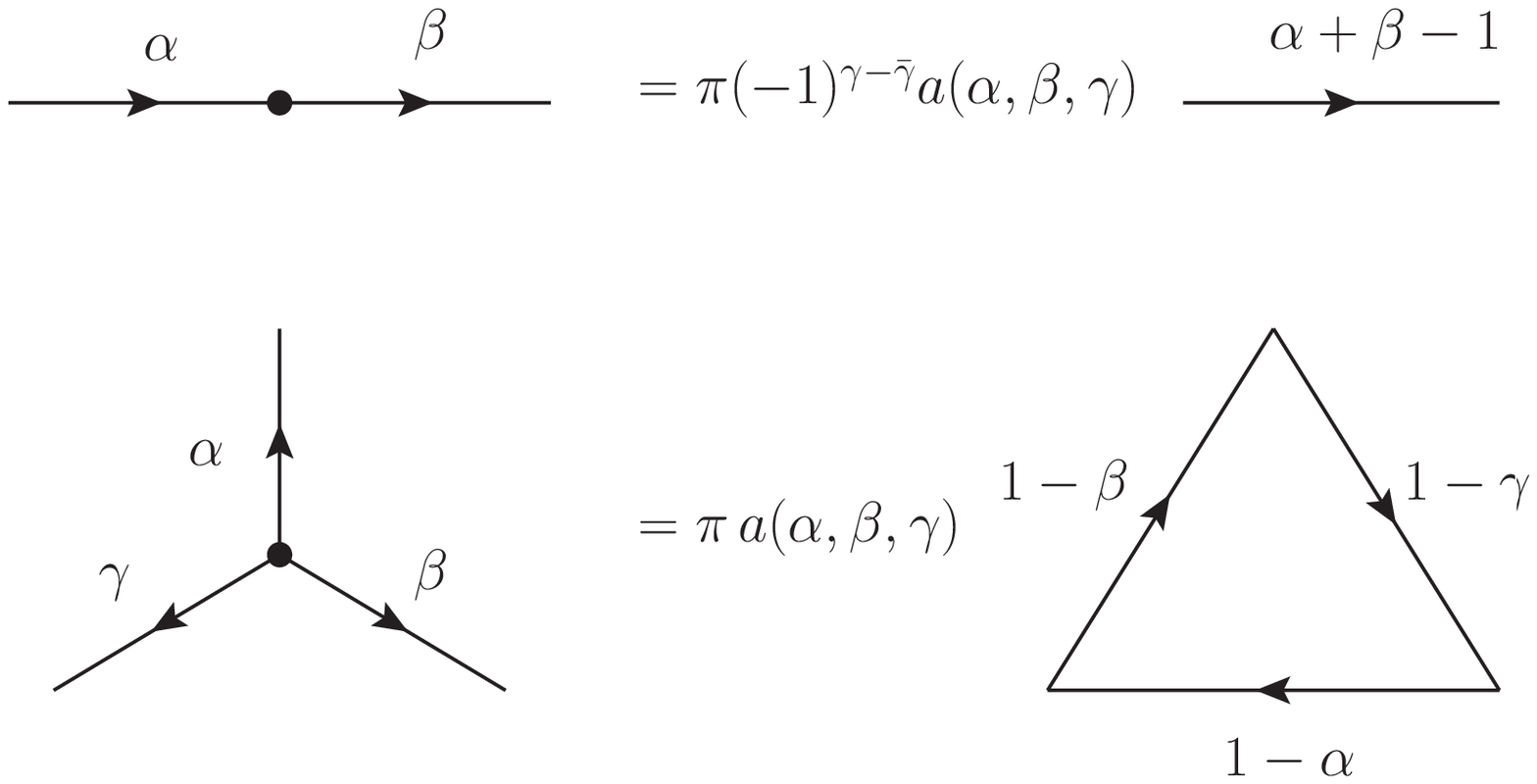}}
\caption{The chain and star-triangle relations, $\alpha+\beta+\gamma=2$.}
\label{Chain+Star}
\end{figure}

After the Fourier transformation we obtain the propagator in the momentum representation
\begin{equation}\label{Fourier}
\int d^2 z \frac{e^{i(pz+\bar p\bar z)}}{[z]^\alpha}=\pi\, i^{\alpha-\bar\alpha}\, a(\alpha)\,\frac{1}{[p]^{1-\alpha}},
\end{equation}
where
\begin{equation}
a(\alpha)=\frac{\Gamma(1-\bar\alpha)}{\Gamma(\alpha)},  \quad a(\bar\alpha)=\frac{\Gamma(1-\alpha)}{\Gamma(\bar\alpha)},
\quad a(\alpha,\beta,\gamma,\ldots):=a(\alpha)a(\beta) a(\gamma)\ldots.
\end{equation}
The function $a(\alpha)$ has the following properties
\begin{eqnarray*} &&
a(\alpha) a(1-\bar\alpha)=1, \qquad \frac{a(1+\alpha)}{a(\alpha)}=-\frac{1}{\alpha\bar\alpha},
\\  &&
a(\alpha)a(1-\alpha)=(-1)^{\alpha-\bar\alpha}, \qquad  a(\alpha)=(-1)^{\alpha-\bar\alpha} a(\bar\alpha).
\end{eqnarray*}

Our evaluations of Feynman diagrams are based on the following computation rules.
\begin{itemize}
\item Chain  relation:
\begin{equation}\label{Chain}
\int d^2 w\frac{1}{[z_1-w]^\alpha [w-z_2]^{\beta}}=
\frac{\pi a(\alpha,\beta,\gamma)}{[z_2-z_1]^{\alpha+\beta-1}},
\end{equation}
where $\gamma=2-\alpha-\beta,\ \bar\gamma=2-\bar\alpha-\bar\beta$.
\item Star-triangle relation:
\begin{equation}\label{Star}
\int d^2w\frac{1}{[z_1-w]^\alpha[z_2-w]^\beta [z_3-w]^\gamma}=
\frac{\pi a(\alpha,\beta,\gamma)}{[z_2-z_1]^{1-\gamma}[z_1-z_3]^{1-\beta}[z_3-z_2]^{1-\alpha}},
\end{equation}
where $\alpha+\beta+\gamma=2$ and $\bar\alpha+\bar\beta+\bar\gamma=2$.
\end{itemize}
These identities are depicted in the diagrammatic form in Fig.~\ref{Chain+Star},
where the blob means the integration over the vertex coordinate.


\end{document}